%% file: IEEE_paper_main.tex
\documentclass[%
	final,
	journal, 
    comsoc,
	letterpaper,
	oneside,
	twocolumn,
	nofonttune,%
]{IEEEtran}%
\input{./organization/preamble.tex}%
\input{./organization/makros.tex}%
\input{./organization/settings.tex}%
\makeatletter
\let\blx@rerun@biber\relax
\makeatother
\usepackage{listings}
\usepackage[usenames,dvipsnames]{xcolor}
\usepackage{tikz} \usetikzlibrary{calc, arrows.meta, intersections, patterns, positioning, shapes.misc, fadings, through,decorations.pathreplacing}

\definecolor{ColorOne}{named}{MidnightBlue}
\definecolor{ColorTwo}{named}{Dandelion}
\definecolor{ColorThree}{named}{Plum}

\usepackage{algorithm}
\usepackage{algpseudocode}
\usepackage{tabularx}
\usepackage{newtxmath}

\usepackage{booktabs}
\pgfplotsset{
  grid style = {
   line width = 0.1pt
  }
}
				\newcommand{\disablewr}[1]{#1}%
				\newcommand{\newcommanddisw}[3]{\newcommand{#1}[1]{\disablewr{\textcolor{#2}{#3}}}}%
\renewcommand{\disablewr}[1]{}%
\definecolor{todocol}{named}{red}
\newcommanddisw{\todo}{todocol}{ToDo: #1}%
\definecolor{migucol}{named}{purple}%
\newcommanddisw{\migucom}{migucol}{{@}comment: #1}%
\newcommanddisw{\miguhigh}{migucol}{#1}%
\begin{document}%
%
\title{%
OpenWiFiSync: Open Source Implementation of a Clock Synchronization Algorithm using Wi-Fi
\thanks{This research was supported by the German Federal Ministry for Digital and Transport (BMDV) within the project 5G-CANKRIN under grant number 19OI23012A. The responsibility for this publication lies with the authors. This is a preprint of a work accepted but not yet published at the 29th IEEE International Conference on Emerging Technologies and Factory Automation (ETFA). Please cite as: M.~Gundall and H.D. Schotten: “OpenWiFiSync: Open Source Implementation of a Clock Synchronization Algorithm using Wi-Fi”. In: 29th IEEE International Conference on Emerging Technologies and Factory Automation (ETFA), IEEE, 2024.}
}
%
\input{./organization/IEEE_authors-long.tex}%
%
%
%
%
%
%
%
\maketitle
%
%
%
%
%
\begin{abstract}%
Precise clock synchronization is an important requirement for distributed and networked industrial use cases. As more and more use cases contain mobile devices, clock synchronization has to be performed over wireless communication links. As wireless communication links are currently not as deterministic and reliable as wireline communication systems, novel clock synchronization algorithms have to be investigated. Here, the so-called Reference Broadcast Infrastructure Synchronization Protocol is a well suited solution as it brings up multiple advantages. Most important is the non-invasiveness, meaning it can be used with commercially available components. As a considerably high amount of factories use \mbox{Wi-Fi} as wireless communication system for their mobile use cases, the aforementioned protocol is implemented using \mbox{Wi-Fi}. Furthermore, the usage of Open-Source Software can be seen as driver for highly efficient and interoperable applications. Consequently, the implementation is accessible under the GNU General Public License on GitHub under the designation OpenWiFiSync. 

Besides the details on concept, its implementation, and the used testbed, first  results are outlined within this paper. Additionally, future work and the estimated timeline are presented.  
\end{abstract}%
\begin{IEEEkeywords}
Clock Synchronization, IEEE 802.11, Wi-Fi, Open-source software
\end{IEEEkeywords}
%
%
%
%
%
\IEEEpeerreviewmaketitle
%
%
%
%
%
%
%
%
\tikzstyle{descript} = [text = black,align=center, minimum height=1.8cm, align=center, outer sep=0pt,font = \footnotesize]
\tikzstyle{activity} =[align=center,outer sep=1pt]

\section{Introduction}%
\label{sec:Introduction}
The 4th Industrial Revolution (I4.0) is fuelled by use cases that rely on an increasing number of mobile Industrial Cyber Physical Systems (ICPS). This leads to a significant shift in the landscape of industrial communication systems, necessitating the use of wireless communications to support novel use cases \cite{7883994}. Here, \gls{5g} is seen as key technology for their realization \cite{8502649}. However, due to several reasons, e.g. pricing, complexity, and regulatory aspects, numerous \mbox{Wi-Fi} installations are currently deployed to fulfill certain industrial mobile use cases \cite{tramarin2015use}. Furthermore, an important requirement in industrial networks is the time alignment across devices. This is where so-called Clock Synchronization Protocols (CSP) come into play, whereby two CSPs in particular, namely the Network Time Protocol (NTP) and the Precision Time Protocol (PTP), have established themselves as de facto standards for wired communication systems due to the required precision and accuracy. For wireless communications, such as \mbox{Wi-Fi}, multiple different CSPs are being investigated~\cite{7782431}. The reason for this is the non-deterministic property of the wireless channel. 

Another emerging trend in research that is currently adopted by industry is the use of so-called Open-Source Software (OSS). This offer numerous advantages, including transparency, efficiency, and reduced vendor lock-in \cite{fuggetta2003open}. In addition, OSS also enables Small and Medium-sized Enterprises (SMEs) to use innovative solutions, even if they do not have the same resources as large companies \cite{colombo2014open}. 

Accordingly, the following contributions can be found in this paper:
\begin{itemize}
  \item Measurements to determine the determinism of different \mbox{Wi-Fi} configurations, i.e. infrastructure and ad hoc mode, as well as Ethernet.
  \item Implementation concept of OpenWiFiSync\footnote{https://github.com/dfki-in-icc/OpenWiFiSync} OSS GitHub project and tentative timeline.
  \item First results of identified clock skew and offset of COTS \mbox{Wi-Fi} adapters. 
\end{itemize}

Accordingly, the paper is structured as follows: Section~\ref{sec:Related Work} gives insights into related work on this topic. The challenges of using clock synchronization algorithms in wireless communications are discussed in Section~\ref{sec:Challenges of Wireless Clock Synchronization}. This is followed by the description of the wireless CSP used in this paper (Section~\ref{sec:rbis}), while Section~\ref{sec:Implementation} details the OpenWiFiSync project. Finally, Section~\ref{sec:Conclusion} concludes the paper.

\section{Related Work}%
\label{sec:Related Work}
Numerous investigations have carried out wireless CSPs. Here, \textit{Mahmood et al.} \cite{7782431} conducted a survey of CSPs that are available for \mbox{Wi-Fi}. Besides \textit{Mahmood et al.}, several other authors, such as \textit{Chen and Yang} \cite{chen2021understanding}, have investigated the use of hardware and software PTP for \mbox{Wi-Fi}. Through the application of major adjustments, a sufficient accuracy and precision can be reached using hardware-based PTP to support most industrial use cases. However, these solutions require dedicated hardware and cannot be used with \gls{cots} \mbox{Wi-Fi} devices. In order to achieve certain degree on precision and accuracy, also other possibilities exist. Here, the usage of the Timing Synchronization Function (TSF) embedded in the IEEE 802.11 standard is motivated by \textit{Mahmood~et~al.}~\cite{6948529}. According to the standard, each Wireless Network Interface Card (WNIC) maintains a 64-bit TSF counter with a 1 MHz tick rate. Hence, microsecond accuracy can theoretically be reached. Furthermore, the properties of wireless channels allow for simultaneous reception of messages at end devices by broadcasting messages. This has lead to investiagtions using one-way message exchange for wireless CSP.
Hence, \textit{Cena et al.} \cite{7018946} introduced the so-called \gls{rbis} protocol. 
Furthermore, this protocol can be used for the integration of \gls{tsn} into \glspl{wlan}~\cite{gundall2021integration}. \textit{Haxhibeqiri et al.} \cite{9453686} are using a similar approach, but motivating the use of TSF counters for distributing timing information. Even if multiple investigations are in the scope of our work, two major concerns arise. Firstly, dedicated hardware is needed, such as Software-Defined Radios (SDRs) or special WNICs with an adopted driver. Secondly, none of the implementations is available as OSS.

\section{Challenges of Wireless Clock Synchronization}%
\begin{figure}[htbp]
\centerline{\includegraphics[width=0.9\columnwidth, trim = {60 180  60  180}, clip]{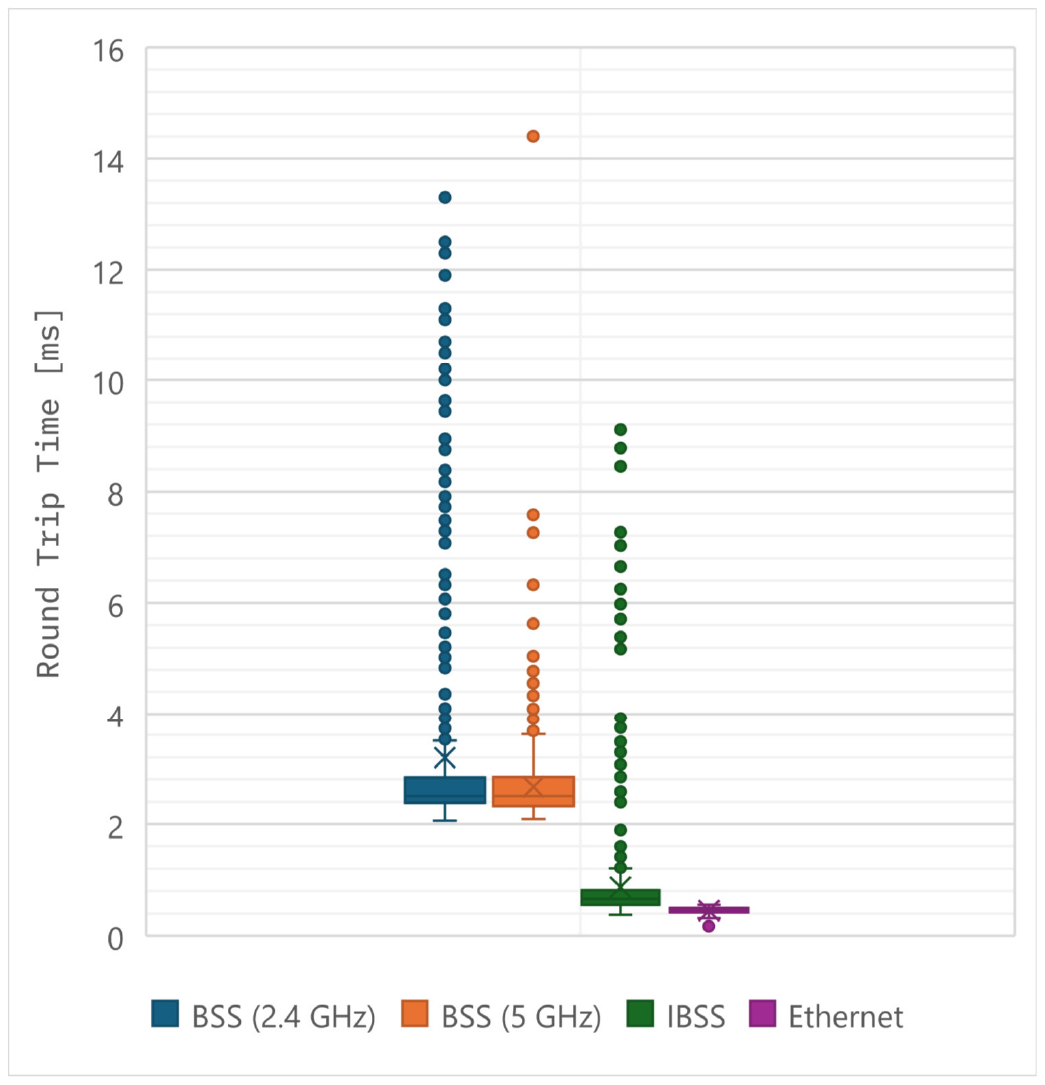}}
\caption{Boxplots of ICMP packets using different communication technologies.}
\label{fig:boxplots}
\end{figure}

\label{sec:Challenges of Wireless Clock Synchronization}
This section outlines the challenges of applying existing wireline CSPs, such as PTP, in wireless communication systems. The major problems are the inherent characteristics of wireless channels that introduce variable transmission delays, leading to timing uncertainties. In order to assess these delays, measurements are performed, using a testbed with the hardware listed in Table~\ref{tab:testbed hardware}.
\begin{table}[htbp]
    \centering
    \caption{Hardware specification.}
    \begin{tabularx}{\columnwidth}{c   c   X}
    \toprule
    \textbf{Equipment} & \textbf{QTY} & \textbf{Specification}\\
    \midrule
    Mini PC & 2 & Intel Core i7-1165G7, 2x16 GB DDR4, Intel Wi-Fi 6 AX201 WNIC, Intel i255-LM Gibgabit NIC, Ubuntu 22.04.4 LTS 64-bit, \linebreak Linux 5.15.0-1063-realtime  \\
    Wi-Fi Router & 1 & Netgear WiFi 6 AX1800 Dual Band Wireless Access Point \linebreak Model: WAX204\\
      \bottomrule 
    \end{tabularx}
    \label{tab:testbed hardware}
\end{table}
Using this testbed that only consists of COTS hardware, a number of 6,000 ICMP packets were exchanged and the \gls{rtt} measured as shown in Figure~\ref{fig:boxplots}. 
Here, several configurations of the \mbox{Wi-Fi} system were applied, i.e. BSS (Basic Service Set) that is also denoted by infrastructure mode and IBSS (Independent Basic Service Set). Latter is also known as ad hoc mode. Furthermore, the \glspl{rtt} of the \mbox{Wi-Fi} tests are compared to Ethernet. Thus, Table~\ref{tab:probabilities} \begin{table}[h]
    \centering
    \caption{Standard deviations of the different communication systems.}
    \begin{tabular}{l   c   c   c}
    \toprule
  \textbf{Comm. system} & $\boldsymbol{\sigma}$~~\textbf{[ms]} & $\mathbf{2\boldsymbol{\sigma}}$~\textbf{[ms]}  & $\mathbf{3}\boldsymbol{\sigma}$~~\textbf{[ms]} \\
     \midrule
  BSS (2.4~GHz)  & 3.19$\pm$1.80 & 3.19$\pm$3.60 & 3.19$\pm$5.39 \\
  BSS (5~GHz)  & 2.67$\pm$0.64 & 2.67$\pm$1.28& 2.67$\pm$1.91 \\
  Ad hoc (2.4~GHz)  & 0.87$\pm$0.93  & 0.87$\pm$1.85  & 0.87$\pm$2.78  \\
  Ethernet & 0.45$\pm$0.05 & 0.45$\pm$0.09 & 0.45$\pm$0.14  \\ 
      \bottomrule 
    \end{tabular}
    \label{tab:probabilities}
\end{table} shows both average values and standard deviations of the measurements. Besides the differences in the averages that could easily be compensated by a CSP, the standard deviations of \mbox{Wi-Fi} significantly differ from Ethernet. While the variation of the RTTs of the Ethernet setup differs from 1~\% to 10~\%, it varies between 23.9~\% (BSS 5 GHz) and  320~\% (IBSS). This leads to a poor performance of CSPs that are applying path delay estimations, such as PTP, without major adoptions.  

\section{Reference Broadcast Infrastructure Synchronization Protocol}%
\begin{figure}[htbp]
\centering
 \subfloat[RBIS protocol.]{\includegraphics[scale=0.95]{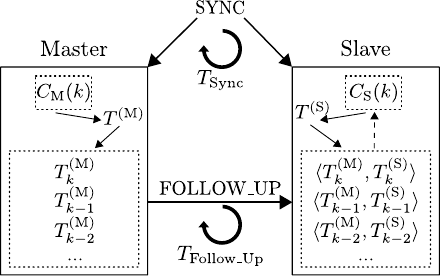}\label{fig:rbis}} 
 
 \subfloat[Sequence diagram of RBIS protocol.]{\includegraphics[scale=0.95]{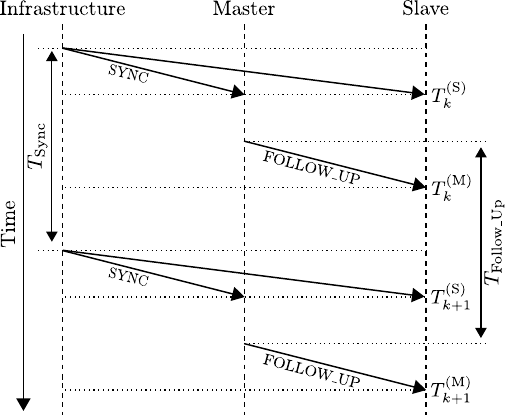}\label{fig:rbis_msc}}
\caption{Visualization of the RBIS protocol (refined from \cite{7018946}).}
\label{fig:rbis protocol}
\end{figure}

\label{sec:rbis}
This section introduces the so-called \gls{rbis} protocol that serves as basis for this paper. As illustrated in Figure~\ref{fig:rbis}, 
the infrastructure transmits a periodic broadcast signal designated as SYNC, which is received by devices within the transmission range. In \mbox{Wi-Fi}, beacon frames can be used as SYNC message. They contain the SSID of the \gls{ap}, the time interval of the transmission, and the timestamp of the beacon, i.e. the time that elapsed since the \gls{ap} was powered. This interval can be adjusted, whereas the default is 102.4 ms. Upon reception of SYNC, each device generates a timestamp based on its individual clock. Subsequently, the master transmits its \grqq correct" timestamps, denoted by $T^\mathrm{M}_k$, to all slaves via the FOLLOW\_UP message. Subsequently, each slave generates its own reference timestamps, denoted by $T^\mathrm{S}_k$. The resulting $T^\mathrm{M}_k$ and $T^\mathrm{S}_k$ tuples are used to estimate the offset $\theta$ and skew $\gamma$ between the master and slave clocks $C_\mathrm{M}(k)$ and $C_\mathrm{S}(k)$ following Equations~\ref{eq:1}-\ref{eq:2}~\cite{6817598}.
\begin{equation}
\hat{\theta}(k)=T^\mathrm{(S)}(k)-{T}^\mathrm{(M)}(k)
\label{eq:1}
\end{equation}
\begin{equation}
\hat{\gamma}(k)=\frac{\hat{\theta}(k)-\hat{\theta}(k-1)}{{T}_\mathrm{M}(k)-{T}_\mathrm{M}(k-1)}
\label{eq:2}
\end{equation}
Figure~\ref{fig:rbis_msc} depicts the sequence diagram of the message exchange. This illustrates that the interval of the SYNC messages does not have to be identical to that of the FOLLOW\_UP messages. 

\section{OpenWiFiSync}%
\label{sec:Implementation}
This section gives insights in the OpenWiFiSync project. Here, the implementation is detailed (see Section~\ref{subsec: 1}). This is followed by preliminary results (see Section~\ref{subsec: 2}). Afterwards, the estimated project timeline is presented (see Section~\ref{subsec: 3}).

\subsection{Implementation}
\label{subsec: 1}
\begin{figure}[htbp]
\centerline{\includegraphics[width=\columnwidth]{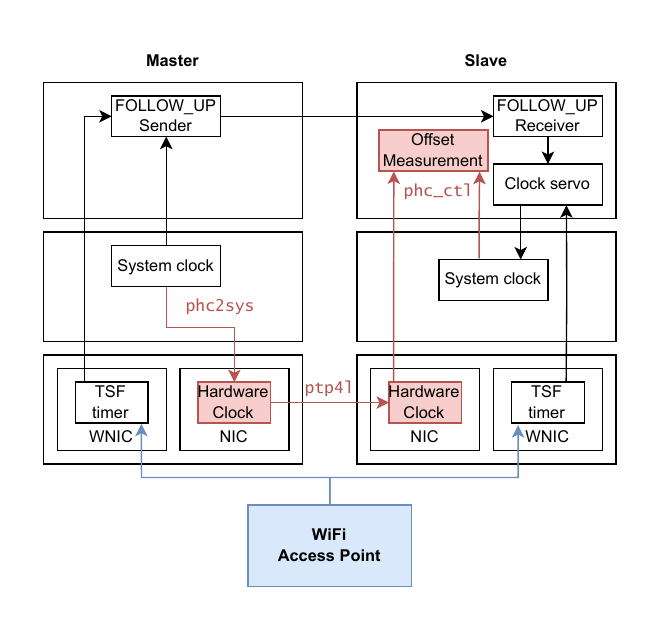}}
\caption{Implementation architecture.}
\label{fig:Software architecture}
\end{figure}

The architecture of the OpenWiFiSync project is shown in Figure~\ref{fig:Software architecture}.
It consists of master and slave. Here, it is assumed that both components have a WNIC including a TSF timer that is synchronized by a WiFi access point that is denoted by the blue color. Furthermore, the master reads both TSF timer and system clock and  transmits this information to the slave. On the other side, the slave receives the information and is able to adopt the system clock accordingly using a clock correction servo. Here, it is planned to integrate one or multiple clock servos implemented in the so-called LinuxPTP project \cite{cochran2015linux}. LinuxPTP a popular OSS project that implements IEEE 1588 PTP protocol and can be used under the GNU General Public License. Hence, similar licence has to be applied to the OpenWiFiSync GitHub project.

Moreover, if both master and slave have a NIC with a hardware clock, which is true for most embedded or PCI(e) NICs, the offset between the system clocks of both devices can be identified. For this, modules provided by the LinuxPTP project can be used to synchronize the hardware clock of the NIC with the system clock (\texttt{phc2sys}) as well as the hardware clocks of different devices (\texttt{ptp4l}). Further, offset measurements between system clock and hardware clocks of a NIC can be performed using \texttt{phc\_ctl} module. Here, the red color indicates the procedure needed to perform the offset measurements.   

\subsection{Preliminary Results}
\label{subsec: 2}
\begin{figure*}[htbp]
\centering
\subfloat[Offset of TSF timestamps of master and slave.]{\includegraphics[width=0.9\columnwidth, trim = {60 50 60 50}, clip]{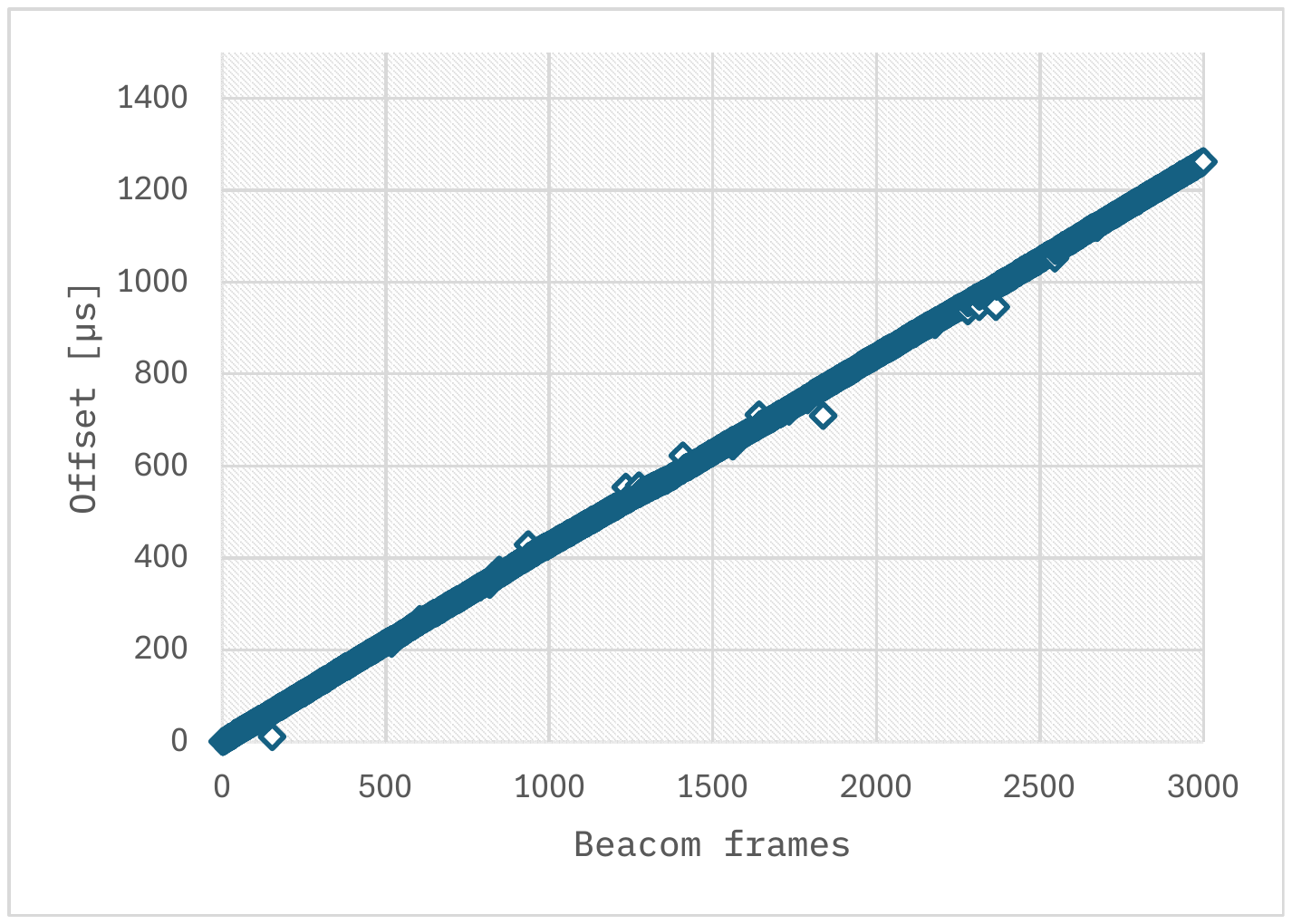}}
\subfloat[Skew of TSF timestamps of master and slave.]{\includegraphics[width=0.9\columnwidth, trim = {60 50 60 50}, clip]{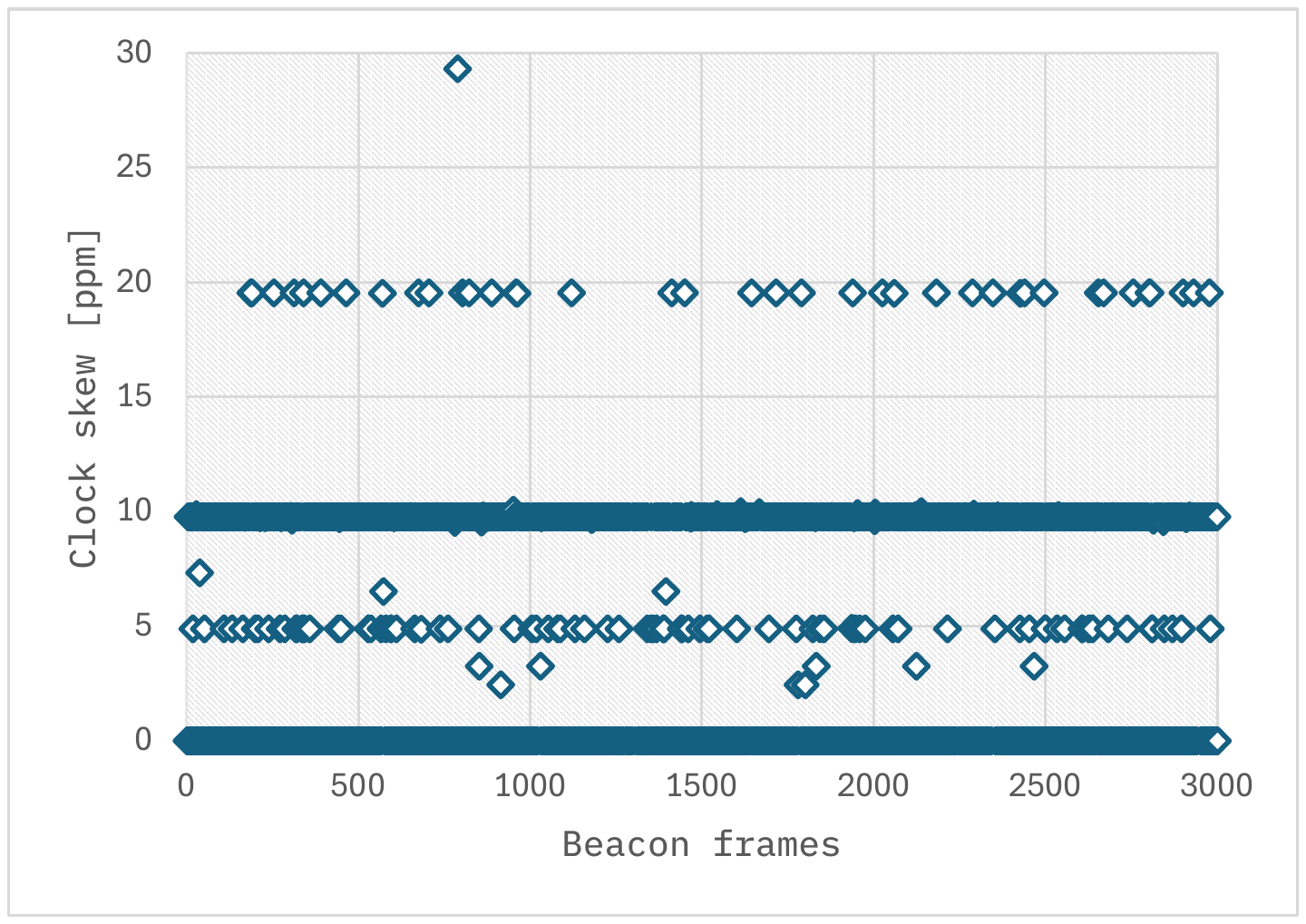}}
\caption{Preliminary results.}
\label{fig:preliminary}
\end{figure*}

In the recent status, the TSF timestamps can be monitored on master as well as slave devices. In addition, both offset and skew can be calculated. Thus, Figure~\ref{fig:preliminary}
depicts offset and skew of 3000 beacon frames ($\approx$ 5 minutes runtime).
The results show a constant skew of the TSF timestamps of master and slave of approximately 3.96~ppm. Furthermore, if a beacon frame is dropped, or the skew identified as zero, for the next frame one or more multiples of the average skew is experienced. This deterministic behaviour can be seen as solid basis for the implementation of the clock servo, whereas one to low two digit accuracy can be assumed. 

\subsection{Project Timeline}
\label{subsec: 3}
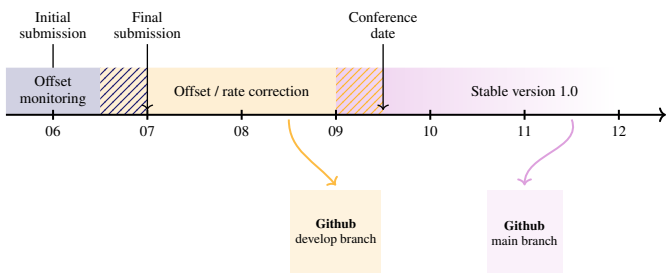
\begin{figure}[htbp]
\resizebox{\columnwidth}{!}{%
\input{figures/timeline.tikz}}
\caption{Tentative timeline of OpenWiFiSync GitHub project.}
\label{fig:timeline}
\end{figure}

As the aforementioned clock correction servos are not yet integrated, the \mbox{OpenWiFiSync} project is still in an early phase. For the different development steps until a stable release is planned, a estimated timeline for the \mbox{OpenWiFiSync} GitHub project is shown in Figure~\ref{fig:timeline}. Here, three major milestones are initially planned to be submitted to GitHub. These are i) offset monitoring, ii) offset/rate correction, and iii) first stable version on the main branch.

\pagebreak
\section{Conclusion}%
\label{sec:Conclusion}
This paper introduces the advantages of OSS, whereas the major advantages are efficiency and interoperability. Furthermore, the challenges for wireless clock synchronization algorithms are outlined. Here, it is shown that the determinism of wireless systems is much lower as for wireline systems. In the case of \mbox{Wi-Fi}, the standard deviation for ICMP packet exchange is in the magnitude of the RTT or even multiple times. Therefore, path delay cannot easily approximated. Hence, a one way protocol is a suitable approach to overcome this issue. Thus, the so-called RBIS protocol is implemented and made publicly available as OSS on GitHub under the project name OpenWiFiSync. First results indicate that one to two digit µs accuracy can be reached using COTS \mbox{Wi-Fi} components without adopting hardware drivers. 
\newpage




\pagebreak
\printbibliography%
\pagebreak

%
%
\end{document}

%% file: organization/preamble.tex
\PassOptionsToPackage{usenames,dvipsnames,svgnames,x11names,table,prologue}{xcolor}%
\PassOptionsToPackage{hyphens}{url}%
\PassOptionsToPackage{nomessages}{fp}%
\ifCLASSINFOpdf
  \usepackage[pdftex]{graphicx}
\else
  \usepackage[dvips]{graphicx}
\fi
\ifCLASSOPTIONcompsoc
   \usepackage[caption=false,font=normalsize,labelfont=sf,textfont=sf]{subfig}
\else
   \usepackage[caption=false,font=footnotesize]{subfig}
\fi

\usepackage{stfloats}
\usepackage[english]{babel}%
\usepackage[utf8]{inputenc}%
\usepackage[babel,style=english]{csquotes}%
\usepackage{hyphsubst}%
\usepackage[%
	activate={true,%
	nocompatibility},%
	final,%
	tracking=true,%
	kerning=true,%
	spacing=true,%
	factor=1100,%
	stretch=10,%
	shrink=10%
]{microtype}%
\usepackage{setspace}%
\usepackage{xcolor}%
\definecolor{todonotecol}{RGB}{250,0,0}%
\usepackage{xparse}
\usepackage[%
	colorlinks=false,%
	urlcolor=black,%
	linkcolor=black,%
	citecolor=black,%
	filecolor=black,%
	breaklinks,%
	]{hyperref}%
\usepackage{url}%
\usepackage[%
	acronym,%
	nopostdot,%
	seeautonumberlist,%
	shortcuts,%
	section=chapter,%
	toc,%
]{glossaries}%
\loadglsentries{./supply/glossaries.tex}\glsdisablehyper
\usepackage[%
	backend=biber,%
	style=ieee,%
	isbn=false,%
	hyperref=true,%
	maxbibnames=99,%
	sorting=none,%
	natbib=true,%
	language=english,%
	defernumbers=true,%
	]{biblatex}%
\DeclareFieldFormat{sentencecase}{\csname bbx@colon@search\endcsname#1}

\usepackage{nameref}
\usepackage{soul}
\usepackage{flushend}
\usepackage{textcomp}
\usepackage{calc}
\usepackage{xkeyval}
\usepackage{multirow}
\usepackage{tabulary}
\usepackage{makecell}
\usepackage{pgfplots}
\usepackage{tikz}
\usepackage{textcomp} 
\usepackage{verbatim}

%% file: supply/glossaries.tex
\glsaddkey*{url}
{}
{\glsentryurl}
{\Glsentryurl}
{\glsurl}
{\Glsurl}
{\GLSurl}
\glsaddkey*{longpltwo}
{}
{\glsentrylongpltwo}
{\Glsentrylongpltwo}
{\glslongpltwo}
{\Glslongpltwo}
{\GLSlongpltwo}
%
%
%
%
%
\newglossaryentry{}{%
	name={},%
	plural={},%
	description={},%
}%
%
%
%
%
%
%
\newacronym{it}{IT}{information technology}
\newacronym{ot}{OT}{operational technology}
\newacronym{opcua}{OPC UA}{Open Platform Communications Unified Architecture}
\newacronym{tsn}{TSN}{Time-Sensitive Networking}
\newacronym{scada}{SCADA}{supervisory control and data acquisition}
\newacronym{vpc}{VPC}{Virtual Process Controller}
\newacronym{plc}{PLC}{Programmable Logic Controller}
\newacronym{udp}{UDP}{User Datagram Protocol}
\newacronym{pubsub}{PubSub}{Publish Subscribe}
\newacronym{vm}{VM}{virtual machine}
\newacronym{pc}{PC}{personal computer}
\newacronym{ptp}{PTP}{Precision Time Protocol}
\newacronym{gptp}{gPTP}{generalized Precision Time Protocol}
\newacronym{ntp}{NTP}{Network Time Protocol}
\newacronym{splc}{Soft-PLC}{Software-PLC}
\newacronym{rte}{RTE}{Real-Time Ethernet}
\newacronym{osi}{OSI}{Open Systems Interconnection}
\newacronym{mac}{MAC}{Media Access Control}
\newacronym{e2e}{E2E}{end-to-end}
\newacronym{tcp}{TCP}{Transmission Control Protocol}
\newacronym{hmi}{HMI}{Human-Machine Interface}
\newacronym{kpi}{KPI}{Key Performance Indicator}
\newacronym{mes}{MES}{Manufacturing Execution System}
\newacronym{rami4.0}{RAMI 4.0}{Reference Architectural Model Industrie 4.0}
\newacronym{tacnet4.0}{TACNET 4.0}{Tactile Internet 4.0}
\newacronym{3gpp}{3GPP}{3rd Generation Partnership Project}
\newacronym{5gppp}{5GPPP}{5G Infrastructure Public Private Partnership}
\newacronym{itu}{ITU}{International Telecommunication Union}
\newacronym{ngnm}{NGNM}{Next Generation Mobile Networks}
\newacronym{agv}{AGV}{automated guided vehicle}
\newacronym{v2v}{V2V}{vehicle-to-vehicle}
\newacronym{v2x}{V2X}{vehicle-to-infrastructure}
\newacronym{d2x}{D2X}{device-to-x}
\newacronym{qos}{QoS}{quality of service}
\newacronym{noa}{NOA}{NAMUR Open Architecture}
\newacronym{dcs}{DCS}{distributed control system}
\newacronym{m2m}{M2M}{machine-to-machine}
\newacronym{sdn}{SDN}{software-defined networking}
\newacronym{erp}{ERP}{enterprise resource planning}
\newacronym{ip}{IP}{Internet Protocol}
\newacronym{lte}{LTE}{Long Term Evolution}
\newacronym{5g}{5G}{5th generation wireless communication systems}
\newacronym{harq}{HARQ}{Hybrid Automatic Repeat Request}
\newacronym{rat}{RAT}{radio access technology}
\newacronym{mc}{MC}{Multi-Connectivity}
\newacronym{fbmc}{FBMC}{Filter Bank Multicarrier}
\newacronym{gfdm}{GFDM}{Generalized Frequency Division Multiplexing}
\newacronym{sla}{SLA}{service-level agreement}
\newacronym{5qi}{5QI}{5G QoS Indicator}
\newacronym{bmbf}{BMBF}{German Federal Ministry of Education and Research}
\newacronym{nrt}{NRT}{non real-time}
\newacronym{irt}{IRT}{isochronous real-time}
\newacronym{rt}{RT}{real-time}
\newacronym{ar}{AR}{augmented reality}
\newacronym{wlan}{WLAN}{Wireless Local Area Network}
\newacronym{cpu}{CPU}{central processing unit}
\newacronym{ram}{RAM}{random-access memory}
\newacronym{ue}{UE}{user equipment}
\newacronym{ran}{RAN}{radio access network}
\newacronym{bs}{BS}{base station}
\newacronym{cn}{CN}{core network}
\newacronym{epc}{EPC}{evolved packet core}
\newacronym{mec}{MEC}{mobile edge cloud}
\newacronym{rtt}{RTT}{Round Trip Time}
\newacronym{vpn}{VPN}{virtual private network}
\newacronym{vpf}{VPF}{Virtual Process Control Function}
\newacronym{vdi}{VDI}{Association of German Engineers}
\newacronym{ipc}{IPC}{industrial pc}
\newacronym{rtos}{RTOS}{real-time operating system}
\newacronym{mano}{VPC M\&O}{VPC Manager \& Orchestrator}
\newacronym{i/o}{I/O}{Input / Output}
\newacronym{fbd}{FBD}{Function Block Diagram}
\newacronym{ldd}{LDD}{Ladder Logic Diagram}
\newacronym{sfc}{SFC}{Sequential Function Chart}
\newacronym{il}{IL}{Instruction List}
\newacronym{st}{ST}{Structured Text}
\newacronym{os}{OS}{Operating System}
\newacronym{mqtt}{MQTT}{message queuing telemetry transport }
\newacronym{amqp}{AMQP}{advanced message queuing protocol}
\newacronym{api}{API}{application programming interface}
\newacronym{embb}{eMBB}{enhanced mobile broadband}
\newacronym{iot}{IoT}{Internet of Things}
\newacronym{urllc}{URLLC}{ultra-reliable low-latency communication}
\newacronym{mmtc}{mMTC}{massive machine-type communications}
\newacronym{iiot}{IIoT}{industrial IoT}
\newacronym{iiot2}{IIoT}{industrial Internet of Things}
\newacronym{npn}{NPN}{non-public network}
\newacronym{prb}{PRB}{Physical Ressource Block}
\newacronym{gnb}{gNB}{5G base station}
\newacronym{5gs}{5GS}{5G system}
\newacronym{5gc}{5GC}{5G core network}
\newacronym{oai}{OAI}{OpenAirInterface}
\newacronym{enb}{eNB}{4G base station}
\newacronym{sdr}{SDR}{Software Defined Radio}
\newacronym{usrp}{USRP}{Universal Software Radio Peripheral}
\newacronym{mme}{MME}{Mobility Management Entity}
\newacronym{hss}{HSS}{Home Subscriber Server}
\newacronym{pdn}{PDN}{Packet Data Network}
\newacronym{sgw}{SGW}{Serving Gateway}
\newacronym{gtp}{GTP}{GPRS Tunneling Protocol}
\newacronym{mimo}{MIMO}{Multiple Input Multiple Output}
\newacronym{fpga}{FPGA}{Field Programmable Gate Array}
\newacronym{pss}{PSS}{primary synchronization signal}
\newacronym{sss}{SSS}{secondary synchronization signal}
\newacronym{sfn}{SFN}{system frame number}
\newacronym{mib}{MIB}{master information block}
\newacronym{pbch}{PBCH}{Physical Broadcast Channel}
\newacronym{ism}{ISM}{Industrial, Scientific and Medical}
\newacronym{e-utra}{E-UTRA}{evolved UMTS Terrestrial Radio Access}
\newacronym{bldc}{BLDC}{brushless DC}
\newacronym{dl}{DL}{Downlink}
\newacronym{ds-tt}{DS-TT}{Device-Side TSN Translator}
\newacronym{nw-tt}{NW-TT}{Network-Side TSN Translator}
\newacronym{upf}{UPF}{User Plane Function}
\newacronym{tsi}{TSi}{ingress timestamping}
\newacronym{tse}{TSe}{egress timestamping}
\newacronym{gm}{GM}{grandmaster}
\newacronym{sn}{SN}{sequence number}
\newacronym{rbis}{RBIS}{Reference Broadcast Infrastructure Synchronization}
\newacronym{cococo}{CoCoCo}{communication-control codesign}
\newacronym{3c}{3C}{joint design of communications, control, and computing}
\newacronym{cots}{COTS}{Commercial Off-The-Shelf}
\newacronym{ie}{IE}{industrial Ethernet}
\newacronym{ap}{AP}{access point}
\newacronym{ofdm}{OFDM}{Orthogonal Frequency-Division Multiplexing}
\newacronym{csma}{CSMA-CA}{carrier sense multiple access with collision avoidance}
\newacronym{dcf}{DCF}{distributed coordination function}
\newacronym{pcf}{PCF}{point coordination function}
\newacronym{hcf}{HCF}{hybrid coordination function}
\newacronym{edca}{EDCA}{enhanced distributed channel access}
\newacronym{hcca}{HCCA}{\gls{hcf} controlled channel access}
\newacronym{tdma}{TDMA}{time division multiple access}
\newacronym{ofdma}{OFDMA}{Orthogonal Frequency-Division Multiple Access}
\newacronym{mumimo}{MU-MIMO}{multi-user \gls{mimo}}
\newacronym{bi}{BI}{beacon interval}
\newacronym{ssid}{SSID}{service set identifier}

%% file: organization/makros.tex
\newcommand{\mytilde}{{\raise.17ex\hbox{$\scriptstyle\mathtt{\sim}$}}}

%% file: organization/settings.tex
\newlength\textheighttemp%
\newlength\textwidthtemp%
\newlength\textheightstd%
\newlength\textwidthstd%
\newlength\textheightold%
\newlength\textwidthold%
\newlength\tempheight%
\newlength\tempwidth%
\SetProtrusion{encoding={*},family={bch},series={*},size={6,7}}
              {1={ ,750},2={ ,500},3={ ,500},4={ ,500},5={ ,500},
               6={ ,500},7={ ,600},8={ ,500},9={ ,500},0={ ,500}}
\SetExtraKerning[unit=space]
    {encoding={*}, family={bch}, series={*}, size={footnotesize,small,normalsize}}
    {\textendash={400,400}, 
     "28={ ,150}, 
     "29={150, }, 
     \textquotedblleft={ ,150}, 
     \textquotedblright={150, }} 
\SetTracking{encoding={*}, shape=sc}{40}

%% file: organization/IEEE_authors-long.tex
\author{%
\IEEEauthorblockN{%
    Michael Gundall\IEEEauthorrefmark{1} %
    and Hans D. Schotten\IEEEauthorrefmark{1}\IEEEauthorrefmark{3} %
    \\%
}%
\IEEEauthorblockA{%
    \IEEEauthorrefmark{1}German Research Center for Artificial Intelligence GmbH (DFKI), Kaiserslautern, Germany \\%
    \IEEEauthorrefmark{3}Department of Electrical and Computer Engineering,  RPTU Kaiserslautern-Landau, Kaiserslautern, Germany %
	\\%
    Email: %
        \{michael.gundall, hans\_dieter.schotten%
        \}@dfki.de
        \\%
}%
}%

%% file: figures/timeline.tikz
\begin{tikzpicture}[very thick, black]
\small

\coordinate (O) at (-1,0); 
\coordinate (P1) at (1,0);
\coordinate (P2) at (6,0);
\coordinate (P3) at (12,0);
\coordinate (F) at (13,0); 
\coordinate (E1) at (0,0); 
\coordinate (E2) at (7,0); 
\coordinate (E3) at (2,0); 

\draw[<-,thick,color=black] ($(E1)+(0,0.1)$) -- ($(E1)+(0,1.5)$) node [above=0pt,align=center,black] {Initial\\submission};

\fill[color=ColorOne!20] rectangle (O) -- (P1) -- ($(P1)+(0,1)$) -- ($(O)+(0,1)$); 
\fill[color=ColorTwo!20] rectangle (P1) -- (P2) -- ($(P2)+(0,1)$) -- ($(P1)+(0,1)$); 
\shade[left color=ColorThree, right color=white] rectangle (P2) -- (P3) -- ($(P3)+(0,1)$) -- ($(P2)+(0,1)$); 
\path [pattern color=ColorOne, pattern=north east lines, line width = 1pt, very thick] rectangle ($(O)+(2.0,0)$) -- ($(O)+(3,0)$) -- ($(O)+(3,1)$) -- ($(O)+(2.0,1)$); 
\path [pattern color=ColorTwo, pattern=north east lines, line width = 1pt, very thick] rectangle ($(O)+(7.0,0)$) -- ($(O)+(8,0)$) -- ($(O)+(8,1)$) -- ($(O)+(7.0,1)$); 

\draw ($(P1)+(-1.0,0.5)$) node[activity,Black] {Offset \\ monitoring};
\draw ($(P2)+(-2,0.5)$) node[activity,Black] {Offset / rate correction};
\draw ($(P3)+(-2,0.5)$)  node[activity, Black] {Stable version 1.0};

\node[descript,fill=ColorTwo!15,text=Black](D2) at ($(P2)+(0,-2.5)$) {%
	\textbf{Github}\\
	develop branch};

\node[descript,fill=ColorThree!15,text=Black](D3) at ($(P3)+(-2,-2.5)$) {%
	\textbf{Github}\\
	main branch};
	
\draw[<-,thick,color=black] ($(E2)+(0,0.1)$) -- ($(E2)+(0,1.5)$) node [above=0pt,align=center,black] {Conference\\date};

\draw[<-,thick,color=black] ($(E3)+(0,0.1)$) -- ($(E3)+(0,1.5)$) node [above=0pt,align=center,black] {Final\\submission};

\path[->,color=ColorTwo] ($(P2)+(-1,-0.1)$) edge [out=-90, in=130]  ($(D2)+(0,1)$);
\path[->,color=ColorThree]($(P3)+(-1,-0.1)$)  edge [out=-70, in=90]  ($(D3)+(0,1)$);

\draw[->] (O) -- (F);
\foreach \x in {0,2,...,12}
\draw(\x cm,3pt) -- (\x cm,-3pt);
\foreach \i \j in {0/06,2/07,4/08,6/09,8/10,10/11,12/12}{
	\draw (\i,0) node[below=3pt] {\j} ;
}

\end{tikzpicture}